\documentclass{elsart}
\usepackage{amsmath,amssymb}
\usepackage{cite}
\usepackage{graphicx}
\usepackage{epsfig}
\usepackage{subfigure}

\begin{document}

\begin{frontmatter}

\begin{flushright}
YITP-SB-10-35 \\ ICCUB-10-061
\end{flushright}

\title{CP Violation from Scatterings with Gauge Bosons in Leptogenesis}
\author[cnyitp]{Chee Sheng Fong},
%\ead{fong@insti.physics.sunysb.edu}
\author[cnyitp,icrea,ub]{M.~C.~Gonzalez-Garcia},
%\ead{concha@insti.physics.sunysb.edu}
\author[ub]{J.~Racker}
%\ead{racker@ecm.ub.es}

\address[cnyitp]{C.N. Yang Institute for Theoretical Physics\\
  State University of New York at Stony Brook\\
  Stony Brook, NY 11794-3840, USA.}
\address[icrea]{Instituci\'o Catalana de Recerca i Estudis Avan\c{c}ats 
(ICREA)}
%and 
%Departament d'Estructura i Constituents de la Mat\`eria and ICC-UB\\
%  Universitat de Barcelona, Diagonal 647, E-08028 Barcelona, Spain.}
\address[ub]{ 
Departament d'Estructura i Constituents de la Mat\`eria and ICC-UB\\
  Universitat de Barcelona, Diagonal 647, E-08028 Barcelona, Spain.}

\begin{abstract}
We present an explicit computation of   
the CP asymmetry in scattering processes 
involving the heavy right-handed neutrinos of the type I seesaw framework 
and the Standard Model gauge bosons. Compared to CP violation in
two--body decays and in scatterings with top quarks there are 
new contributions at one loop in the form of new type of vertex
corrections as well as of box diagrams.
We show that their presence implies that, unlike the CP
asymmetry in scatterings with top quarks, the CP asymmetry in 
scatterings with gauge bosons is different from the two-body
decay asymmetry even for hierarchical right-handed neutrinos.
This also holds for the L-conserving CP asymmetry in scatterings 
with U(1)$_Y$ gauge bosons.
\end{abstract}


\end{frontmatter}

The discovery of neutrino oscillations makes leptogenesis a very
attractive solution to the baryon asymmetry problem
\cite{Fukugita:1986,Davidson:2008}.  In the standard framework 
the tiny neutrino masses are generated via the
(type I) seesaw mechanism
\cite{Minkowski:1977,Gell-Mann:1979,Yanagida:1979,Mohapatra:1980} and
since the new singlet neutral leptons have heavy lepton number (L)
violating Majorana masses they can produce dynamically a lepton asymmetry
through out of equilibrium processes. Eventually, this lepton asymmetry is
partially converted into a baryon asymmetry due to fast sphaleron
processes.

In this framework the CP asymmetry can be generated in the $N_i$
decays~\cite{segre}, {i.e.}  $N_i\to l_j H$ ~\cite{Covi:1996} as well
as in scattering processes of the $N_i$ with particles in the
plasma. In perturbation theory, to lowest order, the CP asymmetry 
in any of these processes  arises from the interference between the tree 
level and one--loop amplitudes.  The CP violating asymmetry in 
scatterings with top quarks has been considered in
Refs.~\cite{Pilaftsis:2004,Pilaftsis:2005b,Abada:2006b, Nardi:2007}.
In
Refs.~\cite{Pilaftsis:2004,Pilaftsis:2005b,Abada:2006b} it was argued
that an approximate equality between the scattering and decay
asymmetries should hold  for hierarchical right-handed
neutrinos (RHNs).  Ref.~\cite{Nardi:2007} presented an explicit computation
of the CP asymmetry in top quarks scatterings (TQS) and directly showed
that this was correct.
It is important to note that there is a one to one correspondence
between one--loop diagrams in decays and TQS. Then, as
discussed in Ref.\cite{Abada:2006b}, the ``factorization'' of a common
CP asymmetry can be easily understood in terms of an effective field
theory in which all but the lightest of the RHNs have been integrated
out and their effect appears in a dim-5 operator
$(H\ell)(H\ell)$. In this approximation only ``bubble-like''
diagrams contribute to the one--loop amplitudes of both decays and TQS.

As for the CP asymmetry in scatterings with gauge bosons, its
effect was estimated in Ref.\cite{Nardi:2007} under the assumption
that it also factorizes in terms of the decay CP asymmetry.  However,
the same argument that leads to the understanding of the equality
between the CP asymmetries in decay and TQS
does not hold for gauge boson scatterings (GBS)   due to the presence of
additional contributions to the amplitude at one loop (which in
the effective field theory approximation contain three rather than two
particles in the loop). With this motivation in this note we carry
out an explicit calculation of the CP asymmetry in GBS 
\footnote{
In principle one must consider all processes at a given order in
the coupling constants. In particular the three body
decays $N_i \to \ell_j H A$ should be included together with the gauge
boson scatterings. However the three body decay rate is phase space
suppressed, thus the processes $N_i \to \ell_j H A$ and their CP asymmetries 
can be safely neglected (see e.g. the appendix of~\cite{Nardi:2007}).}
.

For a given heavy neutrino species (called here $N_i$) there are three different types
of GBS, namely 
\begin{equation}
N_{i}A  \to \ell_{j}H \,,  \,\;\;\;\;\;\;\;\;
N_{i}\overline{\ell}_{j}  \to  HA \, ,\;\;\;\;\;\;\;\;\;
\overline{\ell}_{j}A  \to  N_{i}H \, ,
\label{eq:process}
\end{equation}
and each one has an associated CP conjugate process. 
Here $A=W_a$ or $B_Y$ for SU(2)$_L$ and U(1)$_Y$ bosons, 
respectively. 
The Lagrangian for the Yukawa  and gauge interactions relevant for 
the computation of these processes reads
\begin{eqnarray}
\mathcal{L}_{Y+A} & = & 
%&-Y_t \overline{Q}_{3}P_{R} t \widetilde H 
- Y_{N_{ij}}\overline{\ell}_{j}
P_{R}N_{i} \widetilde H 
- \frac{ig_{2}}{2}\left(\partial_{\mu}H\right)^{\dagger}W_a^{\mu}\sigma_{a} H 
-\frac{ig_{1}}{2}\left(\partial_{\mu}H\right)^{\dagger}B_Y^{\mu}H
+\mbox{h.c.} \nonumber \\
&+&\sum_j \left[\frac{g_{2}}{2}\overline{\ell}_{j}
\gamma_{\mu}W_{a}^{\mu}\sigma_{a}
P_{L}\ell_{ j}-\frac{g_{1}}{2}
\overline{\ell}_{j}\gamma_{\mu}B_Y^{\mu}P_{L}\ell_{j}\right],
\label{eq:lag}
\end{eqnarray}
where $i,j$ are generation indices of RHNs   
and lepton doublets $\ell_{j}^{T}=\left(\nu_{j},\ell_{j}^{-}\right)$ 
respectively, 
$P_{L,R}=\frac{1}{2}\left(1\mp\gamma_{5}\right)$, 
and $\widetilde H=i\tau_2 H^*$ with  
$H^{T}=\left(h^{+},h^{0}\right)$ being the Higgs doublet.
\begin{figure}
\centering\includegraphics[scale=0.37]{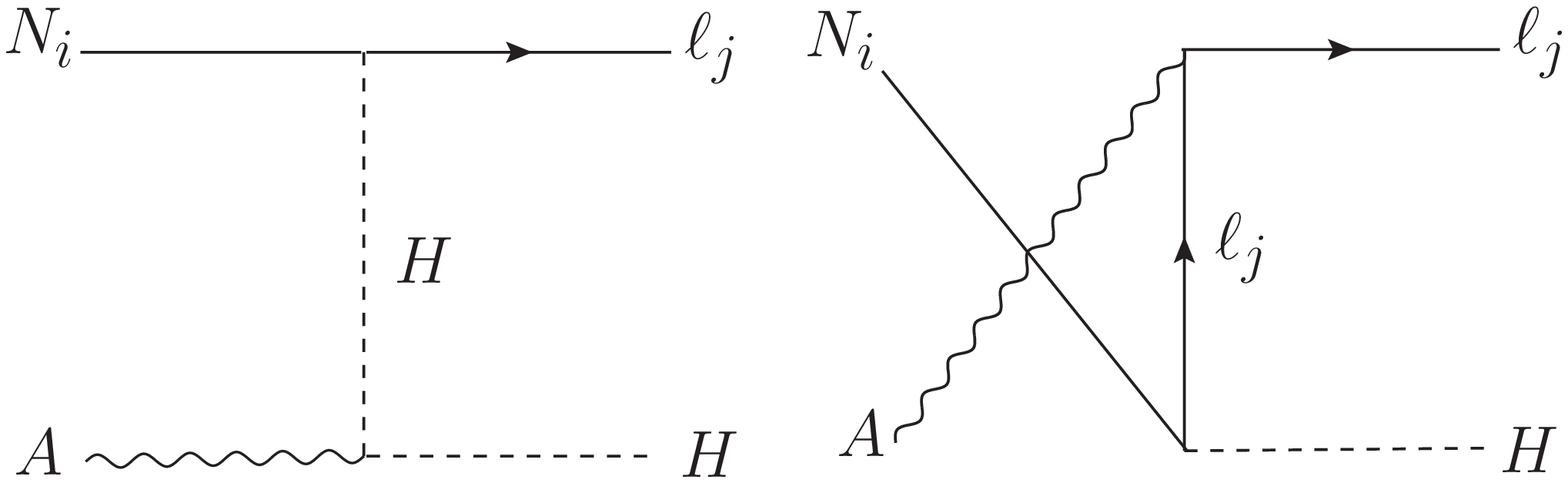}
\centering\includegraphics[scale=0.37]{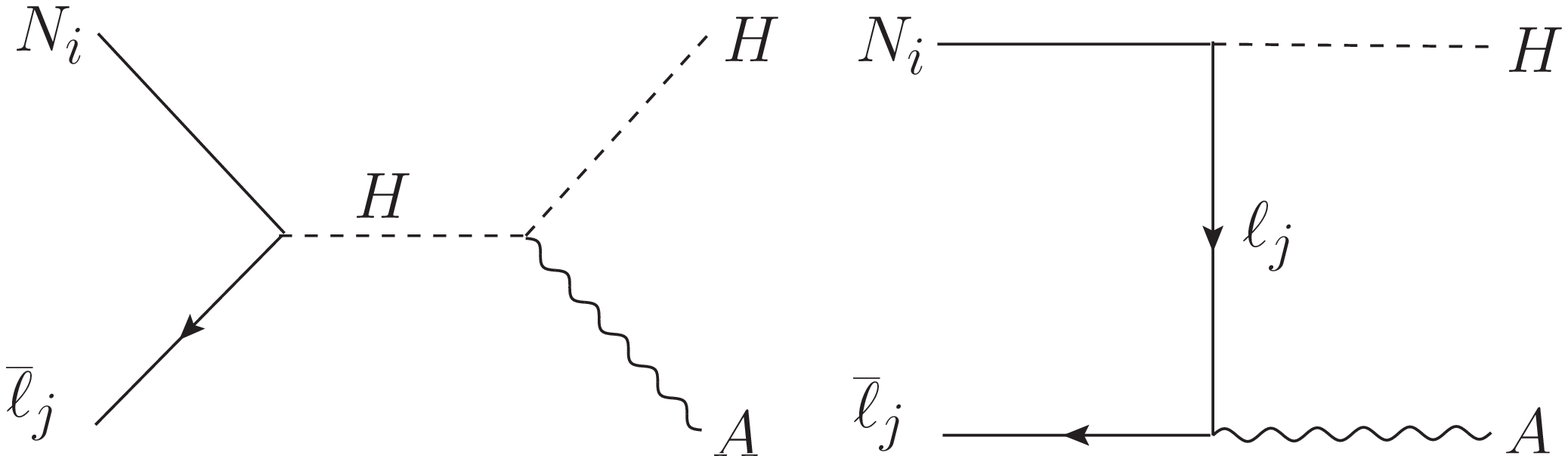}
\centering\includegraphics[scale=0.37]{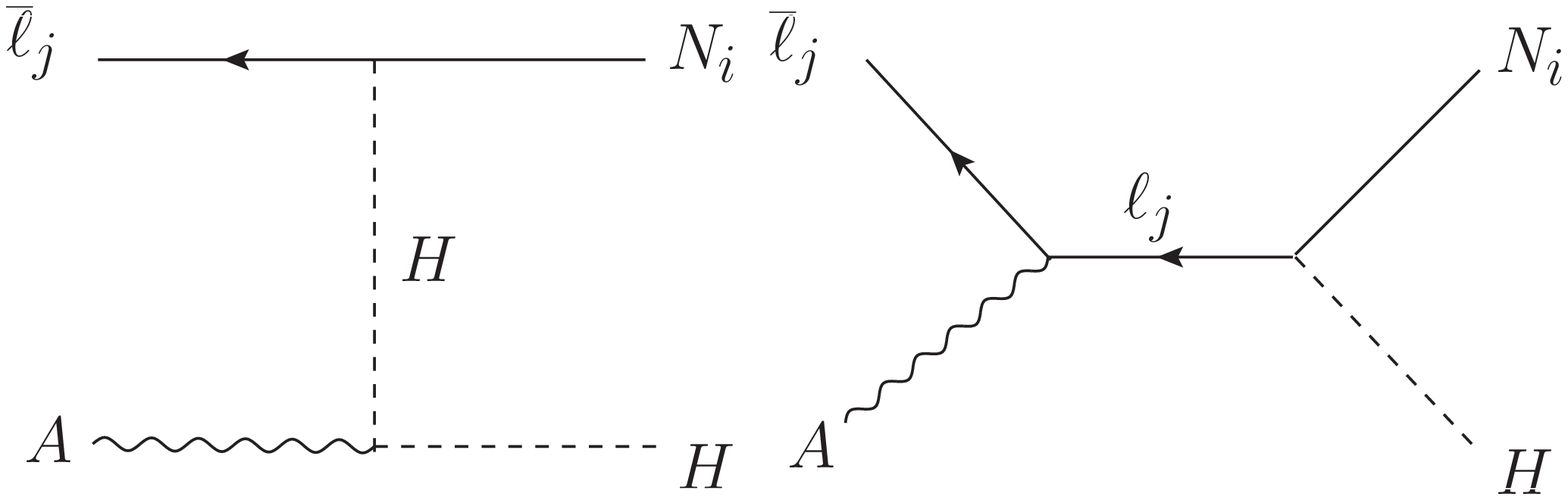}
\caption{Tree diagrams for the three scattering processes 
in Eq.~\eqref{eq:process}.
\label{fig:tree}}
\end{figure}

We denote by $\gamma^A_B\equiv \gamma(A\to B)$ 
the thermally averaged density rate
for a state $A$ to go into the state $B$ (summed over initial
and final spin and gauge degrees of freedom), 
and by $\Delta \gamma^A_B\equiv \gamma^A_B-\gamma^{\bar A}_{\bar B}$ 
the CP difference between the rates for particle and  
antiparticle processes. 
Neglecting the thermal motion with respect to the plasma, 
the relevant thermal average rates for $1\; 2 \rightarrow 3\; 4$ processes can be written as 
\begin{equation}
\gamma^{12}_{34} =
\frac{T}{64 \pi^4} \int_{s_{-}}^\infty ds
\sqrt{s}\; \hat s^{12}_{34}(s)\; K_1\left(\frac{\sqrt{s}}{T}\right) 
%\nonumber \\
%&=& 
%
=\frac{T}{2^9 \pi^5} \int_{s_{-}}^\infty ds\int_{t_-}^{t_+} dt 
\frac{|M^{12}_{34}|^2}{\sqrt{s}}
\; K_1\left(\frac{\sqrt{s}}{T}\right),
\nonumber
\end{equation}
where $\hat s^{12}_{34}(s)$ is the reduced cross section,  
$K_1(x)$ is the modified Bessel function of second kind
of order 1, and $T$ is the temperature. We have also introduced the Mandelstam variables
in terms of the momenta $p_i$ of the four particles in the process,
$s=(p_1+p_2)^2$, $t=(p_1-p_3)^2$, and $u=(p_1-p_4)^2$, with $p_i^2=m_i^2$.
Thus for each process $s_{-}={\rm Max}\left[(m_1+m_2)^2,(m_3+m_4)^2\right]$
and   
\begin{equation}
t_{\mp}={\small \frac{(m_1^2\!-\!m_2^2\!-\!m_3^2\!+\!m_4^2\!)^2}{4s}\! -\!
\left[\! \sqrt{\frac{(s\!+\!m_1^2\!-\!m_2^2)^2}{4s}\!-\!m_1^2}
\!\pm\! \sqrt{\frac{(s\!+\!m_3^2\!-\!m_4^2)^2}{4s}\!-\!m_3^2}\right]^2 }.
\nonumber
\end{equation}

For the sake of simplicity in our evaluation of the CP
asymmetries we neglect the thermal mass of the gauge bosons, and we 
include the thermal masses of leptons $m_\ell$ and the Higgs boson
$m_H$ only in the propagators (but not in the loops), in order to
regularize the infrared divergences that appear when these states are
exchanged in a $t$- or $u$-channel.

In this approximation the tree level amplitudes for the three processes
in \eqref{eq:process}  
are related by crossing symmetry and are given by
\begin{eqnarray}
|M(N_{i}A  \to  \ell_{j}H)|^2=|M(N_{i}A  \to  \bar\ell_{j}\bar H)|^2
&=&2\, {\mathcal G}_A \,|Y_{N_{ij}}|^2 I_{0}(s,t,u), 
\label{eq:tree1}\\
|M(N_{i}\overline{\ell}_{j}  \to  HA)|^2 =|M(N_{i}{\ell}_{j}  \to  \bar HA)|^2.
&=&-2\, {\mathcal G}_A\, |Y_{N_{ij}}|^2 I_{0}(u,s,t), 
\label{eq:tree2}\\
|M(\overline{\ell}_{j}A\to  N_{i}H)|^2 =|M({\ell}_{j}A\to  N_{i}\bar H|)^2 &=& 
2\, {\mathcal G}_A \, |Y_{N_{ij}}|^2 I_{0}(u,t,s),
\label{eq:tree3}
\end{eqnarray}
where ${\mathcal G}_A= 3 g_2^2/2$ or $g_1^2/2$ for SU(2)$_L$ and U(1)$_Y$ GBS, respectively.  \\
$I_0(s,t,u)= 2D_{tH}D_{u\ell}sM_i^{2}+D_{u\ell}^{2}u\left(M_i^{2}-s\right)$, and 
$D_{XP}$ comes from the propagator of particle $P$ in the $X$-channel,
$D_{XP} =  (X-m_P^2)^{-1}$. 

\begin{figure}
\centering\includegraphics[width=\textwidth]{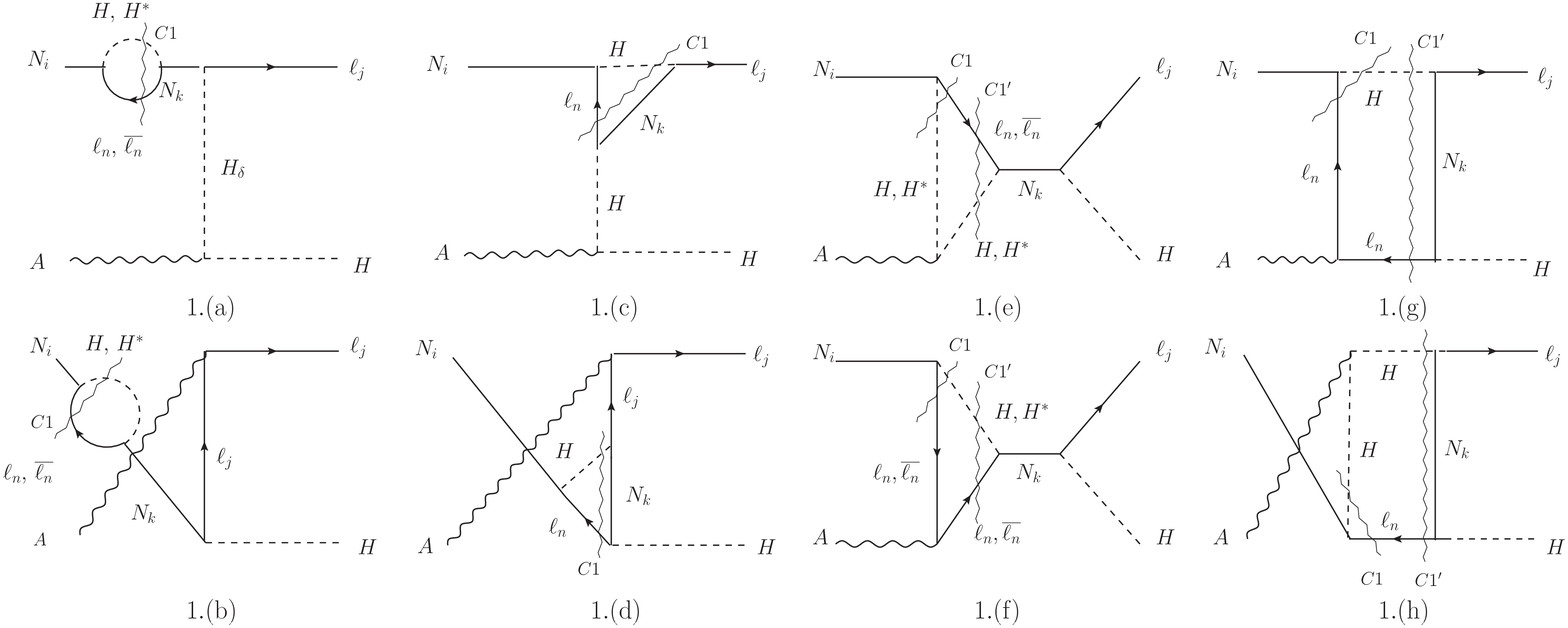}
\centering\includegraphics[width=\textwidth]{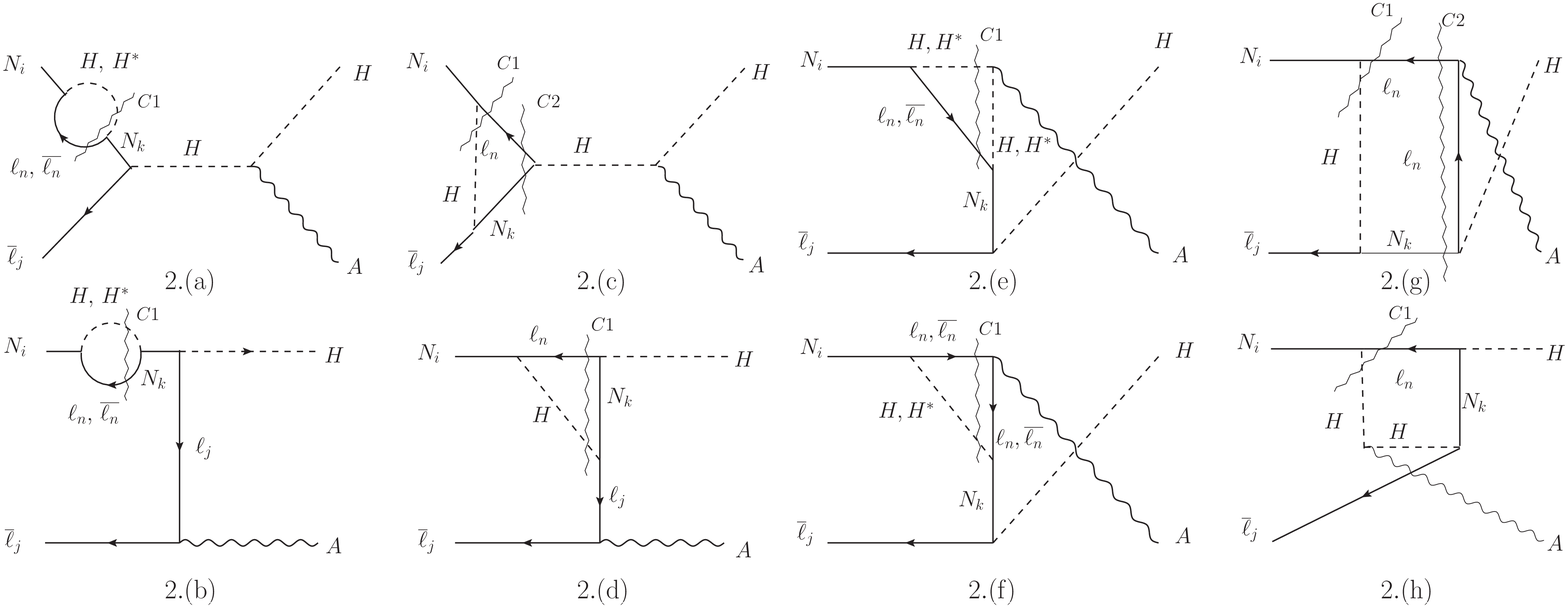}
\centering\includegraphics[width=\textwidth]{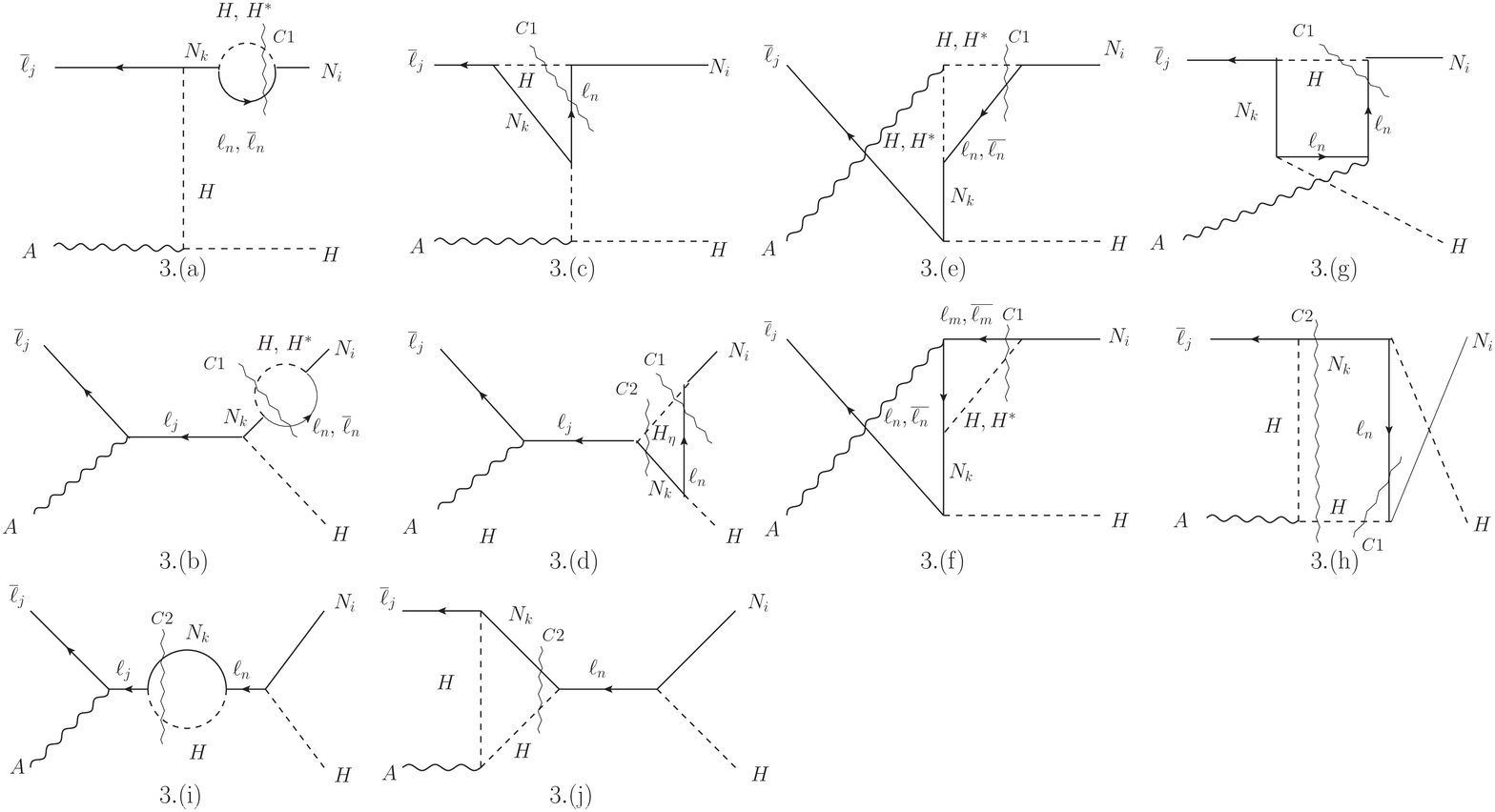}
\caption{One--loop diagrams for the three scattering processes in 
\eqref{eq:process}. We mark with a waving
line the possible branch cuts which contribute to the imaginary part 
of the corresponding amplitude. 
\label{fig:loop}}
\end{figure}

The CP asymmetry in GBS arises from the interference between the
tree-level and one--loop amplitudes in Figs.~\ref{fig:tree}
and~\ref{fig:loop}, respectively.  In Fig.~\ref{fig:loop} we show as a
``waving'' line labeled $C_1$, $C'_1$ or $C_2$ the possible branch
cuts in which the particles in the propagators can be on--shell and
therefore give a contribution to the imaginary part of the
corresponding amplitude.
As seen in Fig.~\ref{fig:loop} besides the $N_i$ self-energy (wave) corrections
(diagrams (a) and (b)), and the vertex corrections
(diagrams (c) and (d)), 
which are also present at one loop in two--body $N_{i}$ 
decays and in TQS, there are new type of vertex diagrams (labeled (e) and
(f)),  as well as contributions from 
boxes (diagrams (g) and (h)).
Furthermore for process  $\overline{\ell}_{j}A  \to  N_{i}H $  
additional imaginary parts appear from 
the last two diagrams 3.(i) and 3.(j).  

For each process the sum of amplitudes of wave diagrams (a) and (b) 
is gauge invariant. The sum of amplitudes for vertex diagrams
(e) and (f) is also gauge invariant, 
while the amplitudes of vertices (c), (d)  and
boxes (g), (h) are not separately gauge invariant but the sum of
the four amplitudes is. Finally for process  
$\overline{\ell}_{j}A  \to  N_{i}H $  
the sum of amplitudes
3.(i) and 3.(j) is gauge invariant.  

Thus generically the CP asymmetry from any of the three processes
in~\eqref{eq:process} 
can be written as
\begin{equation}
{\epsilon^{12}_{34}}\equiv 
\frac{\gamma^{12}_{34}-\gamma^{\bar 1\bar 2}_{\bar 3\bar 4}}
{\displaystyle \sum_j (\gamma^{12}_{34}+\gamma^{\bar 1\bar 2}_{\bar 3\bar 4})}
\equiv{\epsilon^{12}_{34}}^{(w)}
+{\epsilon^{12}_{34}}^{(v1)}+{\epsilon^{12}_{34}}^{(v2)}
+{\epsilon^{12}_{34}}^{(b)}
+\Delta {\epsilon^{\bar\ell_j A}_{N_i H}},
\end{equation}
where ${\epsilon^A_B}^{(w,v1,v2,b)}$ is generated by the
interference between the tree-level amplitudes and 
the self-energy amplitudes (a) and (b), 
the vertex amplitudes (c) and (d), the vertex amplitudes (e) and (f), 
and the box amplitudes (g) and (h), 
respectively. We denote by $\Delta {\epsilon^{\bar\ell_j A}_{N_i H}}$
the extra asymmetry for the process $\overline{\ell}_{j}A  \to  N_{i}H $  
from the diagrams 3.(i) and 3.(j).
Notice that each of the diagrams (a), (b), (e) and (f) represents two amplitudes
depending on whether the internal lepton line is a lepton or an anti-lepton,
which leads respectively to total L-conserving or L-violating contributions.

It is straightforward to show that for any of the processes 
in~\eqref{eq:process} we have
\begin{eqnarray}
&&
{\epsilon^{N_i A}_{\ell_j H}}^{(w)}\!\!
\!\!={\epsilon^{N_i\bar\ell_j}_{A H}}^{(w)}
\!\!={\epsilon^{\bar\ell_j A}_{N_i H}}^{(w)}
\!\!={\epsilon^{N_i}_{\ell j H}}^{(w)}
%\nonumber \\
\!\!=
\frac{\displaystyle\sum_{k\neq i}\left[
C_{jk}\frac{\sqrt{a_k}}{a_k-1}+\widetilde C_{jk}\frac{1}{a_k-1}\right]}
{8\pi\left(Y_{N}Y_{N}^{\dagger}\right)_{ii}},
\label{eq:factorw}\\
%\end{equation}
%where $
&&
\!\!\!\!\!\!\!\!\!\!\!
a_k=\frac{M_k^2}{M^2_i}\,, \;\;
C_{jk}=-\mbox{Im}\left[Y_{N_{ij}}^{*}Y_{N_{kj}}
(Y_{N}Y_{N}^{\dagger})_{ki}\right]\, ,\;\;
%$ and 
%$
\widetilde C_{jk}=
-\mbox{Im}\left[Y_{N_{ij}}^{*}Y_{N_{kj}}(Y_{N}Y_{N}^{\dagger})_{ik}
\right].
\nonumber
%$. \\ 
\end{eqnarray}
In the above, ${\epsilon^{N_i}_{\ell j H}}^{(w)}$ is the contribution to the
CP asymmetry in $N_{i}\to \ell_{j}H$ from the $N_i$ self-energy 
diagrams~\cite{Covi:1996} and 
the first (second) term arises from the L-violating (L-conserving) diagrams.

The contributions to the CP asymmetries from the vertex 
(labeled with superscripts $(v1),(v2)$), box (superscript $(b)$),
and  the extra piece from diagrams 3.(i) and 3.(j)  (superscript $(\Delta)$) 
read
\footnote{When evaluating ${\epsilon^{12}_{34}}^{(v1)}$ and
${\epsilon^{12}_{34}}^{(b)}$ one must specify a gauge since, as
mentioned above, the one--loop vertex $(v1)$ and box amplitudes are not
separately gauge-invariant.  In what follows we will work in the
unitary gauge. }
\footnote{Since the resulting expressions are somewhat cumbersome, to
double check our results   
we have computed the asymmetries both by explicit evaluation of the
corresponding loop integrals as well as by using the Cutkosky rules that give
directly the absorptive part of the Feynman diagrams.}  
\begin{eqnarray}
&&\!\!\!\!\!\!\!\!\!\!\!
\left[|M(12\rightarrow 34)|^2-|M(\bar1\bar2\rightarrow\bar3\bar4)|^2\right]
^{(v1),(v2),(b),(\Delta)}= 
-\sum_{k\neq i}\Bigg\{
\frac{{\mathcal G}_A C_{jk} M_{k}M_{i}}{2\pi}
\nonumber \\
&&\times\left[ 
{I^{12}_{34}}^{(v1),(b)}(s,t,u)+{{I'}^{12}_{34}}^{(b)}(s,t,u)
+
\theta(s-M_k^2) 
{J^{12}_{34}}^{(v1),(b)}(s,t,u)\right]\nonumber \\
&&
\!\!\!\!\!\!\!\!\!\!\!
+\frac{{\mathcal F}_A C_{jk} M_{k}M_{i}}{2\pi}
\left[ 
{K^{12}_{34}}^{(v2)}(s,t,u)+{{K'}^{12}_{34}}^{(v2)}(s,t,u)\right] 
+\frac{\widetilde C_{jk} M^2_{i}}{2\pi} \label{eq:verboxcp}\\
&&
\!\!\!\!\!\!\!\!\!\!\!
\times \left[{\mathcal F}_A \left(
\widetilde {{K}^{12}_{34}}^{(v2)}(s,t,u)+\widetilde{{K'}^{12}_{34}}^{(v2)}(s,t,u)\right )
+{\mathcal G}_A\theta(s-M_k^2) 
{J^{12}_{34}}^{(\Delta)}(s,t,u)\right]\Bigg\},
\nonumber
\end{eqnarray}
where ${I^{12}_{34}}^{(v1),(b)}(s,t,u) \equiv {I^{12}_{34}}^{(v1)}(s,t,u) 
+ {I^{12}_{34}}^{(b)}(s,t,u)$  
comprises the contribution from the
$C_1$ cuts of the vertex diagrams (c) and (d) 
(given by ${I^{12}_{34}}^{(v1)}(s,t,u)$), 
and box diagrams (g) and (h) (given by ${I^{12}_{34}}^{(b)}(s,t,u)$).
${{I'}^{12}_{34}}^{(b)}(s,t,u)$ contains the contribution 
from the $C_1^\prime$ cuts present only in the box
diagrams of $N_{i}A\to \ell_{j}H$.  
${J^{12}_{34}}^{(v1),(b)}(s,t,u)$ has the contribution from the
$C_2$ cuts, which are kinematically allowed when 
$s>M_k^{2}$ and they are possible only
in the processes $N_{i}\overline{\ell}_{j}\to HA$
and $\overline{\ell}_{j}A\to N_{i}H$. All these contributions are
L-violating and they are present for both scatterings with SU(2)$_L$ and
U(1)$_Y$ gauge bosons.  
${J^{12}_{34}}^{(\Delta)}(s,t,u)$ has the -- $L$-conserving --  
contribution from the $C_2$ cuts in the diagrams 3.(i) and 3.(j) present 
only for $\overline{\ell}_{j}A\to N_{i}H$.

${K^{12}_{34}}^{(v2)}(s,t,u)$ ($\widetilde {K^{12}_{34}}^{(v2)}(s,t,u)$) 
contains  the contribution from the
$C_1$ cuts of the 
L-violating (L-conserving) 
vertex diagrams in graphs  (e) and (f). 
Similarly, ${{K'}^{12}_{34}}^{(v2)}(s,t,u)$ ($\widetilde
{{K'}^{12}_{34}}^{(v2)}(s,t,u)$) involves the contribution from the
$C_1^\prime$ cuts of the L-violating (L-conserving) diagrams in the
graphs (e) and (f) of process $N_{i}A\to \ell_{j}H$. Notice that
the diagrams (e) and (f) correspond to amplitudes where the gauge
boson is coupled to the one--loop self energy $N_i-N_k$. Since the
RHNs are SU(2)$_L$ singlets, the sum over the
SU(2)$_L$ degrees of freedom cancel in each of these graphs.
This, however, is not the case for the U(1)$_Y$ gauge boson.
Correspondingly the coupling factors ${\mathcal F}_A=0$ 
for $A=W_a$ while ${\mathcal F}_A={\mathcal G_A}={g_1^2}/2$ for $A=B_Y$.

The $C_1$ cuts are kinematically allowed for all
processes and their corresponding contributions to the amplitudes 
are related by crossing symmetry. 
For the amplitudes of diagrams
(c) and (d)  this reads 
${I^{N_i A}_{\ell_j H}}^{(v1)}(s,t,u)=I^{(v1)}(s,t,u)$, 
${I^{N_i\bar\ell_j}_{A H}}^{(v1)}(s,t,u)=-I^{(v1)}(u,s,t)$, and 
${I^{\bar\ell_j A}_{N_i H}}^{(v1)}(s,t,u)=I^{(v1)}(u,t,s)$, 
with
\begin{eqnarray}
&&\!\!\!\!\!\!\!\!\!\!\!\!\!
I^{(v1)}(s,t,u)= \label{eq:vertex1}\\
&&
\!\!\!\!\!\!\!\!\!\!\!\!\!
D_{u\ell}^{2}M_i^2\left[
\frac{\tilde s \tilde u^{2}}
{(1-\tilde u)^{2}}
\left(-1+\frac{a_k}{1-\tilde u}L_{2}(\tilde u)\right)
+\frac{\tilde t\tilde u}{(1-\tilde u)^{2}}\left(1-\frac{a_k+1-\tilde u}{1-\tilde u}L_{2}(\tilde u)\right)
\right]
\nonumber\\&&
\!\!\!\!\!\!\!\!\!\!\!\!\!
+ D_{tH}D_{u\ell}M_i^2\left[
\frac{\tilde s}{1-\tilde t}\left(1-\frac{a_k+1-\tilde t}{1-\tilde t}
L_{2}(\tilde t)\right)
+\frac{\tilde s}{1-\tilde u}
\left(1-\frac{a_k+1-\tilde u}{1-\tilde u}L_{2}(\tilde u)\right)\right]. 
\nonumber
 \end{eqnarray}
The $C_2$  contributions to the vertex asymmetry from diagrams (c) and (d) are
\begin{eqnarray}
{J^{N_i\bar\ell_j}_{A H}}^{(v1)}
(s,t,u)&=&
D_{sH}D_{t\ell} M_i^2 \frac{\tilde u}{1-\tilde s}\left[1-\frac{a_k}{\tilde s}
-\frac{a_k+1-\tilde s}{1-\tilde s}L_{4}(\tilde s)\right],
\nonumber 
\\
{J^{\bar\ell_j A}_{N_i H}}^{(v1)}(s,t,u)
& = & 
D_{tH}D_{s\ell}M_i^2 \frac{\tilde u}{1-\tilde s}\left[a_k-\tilde s
+\frac{a_k+1-\tilde s}{1-\tilde s}L_{4}\left(\tilde s\right)\right] \nonumber 
\\ && 
\!\!\!\!\!\!\!\!\!\!\!\!\!\!\!\!\!\!\!\!\!\!\!\!\!\!\!\!\!\!\!\!\!\!\!\!\!\! + D_{s\ell}^2 M_i^2\frac{\tilde s}{(1-\tilde s)^2}\left[ (\tilde t - \tilde u)(a_k-\tilde s)
+\frac{\tilde t (a_k+1-\tilde s)- a_k\tilde s \tilde u}{1-\tilde s}L_{4}(\tilde s)\right], 
\label{eq:vertex2}
\end{eqnarray}
and ${J^{N_i A}_{\ell_j H}}^{(v1)}(s,t,u)=0$. 
In Eqs.\eqref{eq:vertex1}--\eqref{eq:vertex2} and the following  we define
$\tilde s=\frac{s}{M_i^2}$, $\tilde t=\frac{t}{M_i^2}$, 
$\tilde u=\frac{u}{M_i^2}$, and 
\begin{eqnarray}
&&
L_{1}(x)\equiv \ln\frac{a_k+x}{x},\;\;
L_{2}(x) \equiv  \ln\frac{a_k+1-x}{a_k},\;\;
L_{4}(x) \equiv 
\ln\frac{x(a_k+1-x)}{a_k},
\nonumber \\
&&L_{3}(x,y) \equiv \ln\frac{xy+a_k(1-x)}
{a_k(1-x)}, \;\;  \;\;
L_{5}(x,y)  \equiv  \ln\frac{xy+a_k(1-x)}{a_k}.
\label{eq:def_log}
\end{eqnarray}

For the $C_1$  contributions to the asymmetry from diagrams (e) and (f)  the crossing symmetry reads 
${K^{N_i A}_{\ell_j H}}^{(v2)}(s,t,u)=K^{(v2)}(s,t,u)$, 
${K^{N_i\bar\ell_j}_{A H}}^{(v2)}(s,t,u)=-K^{(v2)}(u,s,t)$, 
${K^{\bar\ell_j A}_{N_i H}}^{(v2)}(s,t,u)=K^{(v2)}(u,t,s)$, and 
equivalently for the $\widetilde K$ functions, with 
\begin{eqnarray}
\!\!\!\!\!\!\!\!\!\!\!\!
K^{(v2)}(s,t,u) & = & \frac{2 \tilde s
\left\{D_{tH}\tilde t\left[(\tilde t-2\tilde u)+\tilde u(1-\tilde s)\right] 
-D_{u\ell}\tilde u\left[3 \tilde t+\tilde u(1-\tilde s)\right] 
\right\}}{(\tilde s-a_k)(\tilde t+\tilde u)^3}, \nonumber \\
\!\!\!\!\!\!\!\!\!\!\!\!
\widetilde K^{(v2)}(s,t,u) & = & \frac{2 \tilde s\tilde u
\left\{D_{tH}\tilde t\left[3-2(\tilde t+\tilde u)\right]
+D_{u\ell}\left[(2\tilde t-\tilde u)-\tilde t^2+\tilde u^2\right] 
\right\}}{(\tilde s-a_k)(\tilde t+\tilde u)^3}.
\label{eq:vertexnew}
\end{eqnarray}
The contribution from the $C'_1$ cuts in diagrams (e) and (f) 
(which are only present for $N_{i}A\to \ell_{j}H$) is  
\begin{equation}
{{K'}^{N_i A}_{\ell_j H}}^{(v2)}\!\!\!\!
(s,t,u)=-K^{(v2)}(s,t,u) \; , 
\;\;\;\; \;
\widetilde{{K'}^{N_i A}_{\ell_j H}}^{(v2)}\!\!\!\!(s,t,u)
=-{\widetilde K}^{(v2)}(s,t,u) \;, 
\end{equation}
and hence the contributions from vertices (e) and (f) to $N_{i}A\to \ell_{j}H$ 
cancel exactly. 

For the box diagrams:
${I^{N_i A}_{\ell_j H}}^{(b)}(s,t,u)=I^{(b)}(s,t,u)\equiv
I_{b1}(s,t,u) +I_{b2}(s,t,u)$, 
${I^{N_i\bar\ell_j}_{A H}}^{(b)}(s,t,u)=-I^{(b)}(u,s,t)$, and 
${I^{\bar\ell_j A}_{N_i H}}^{(b)}(s,t,u)=I^{(b)}(u,t,s)$, 
with 
\vspace{-2mm}
\begin{eqnarray} 
&&
\!\!\!\!\!\!\!\!\!\!\!\!\!
I_{b1}\left(s,t,u\right) =  \nonumber \\
&&
\!\!\!\!\!\!\!\!\!\!\!\!\!
D_{tH}\left\{ \frac{\tilde s\tilde u}{(1-\tilde s)^{2}}-
\frac{\tilde s (a_k+1-\tilde u)}{\tilde u(1-\tilde t)}
L_{2}(\tilde t)
%\right.
%\label{eq:box1_cut1} 
%&  & \left.
+\left[\frac{a_k+\tilde s}{\tilde u}+
\frac{a_k(a_k+1-\tilde t)}
{\tilde s\tilde t+a_k(1-\tilde s)}\right]L_{3}(\tilde s,\tilde t)\right\} 
\nonumber \\
&& 
\!\!\!\!\!\!\!\!\!\!\!\!\!
+D_{u\ell}\left[-1-\frac{\tilde s\tilde t}{(1-\tilde s)^{2}}
+\frac{a_k+\tilde s}{1-\tilde t}L_{2}(\tilde t)
+\frac{(a_k+\tilde s)(a_k+1-\tilde t)}
{\tilde s\tilde t+a_k(1-\tilde s)}L_{3}(\tilde s,\tilde t)\right],
\label{eq:box1_cut1} \\
%\end{eqnarray}
%\begin{eqnarray}
&&
\!\!\!\!\!\!\!\!\!\!\!\!\!
I_{b2}\left(s,t,u\right) = \nonumber \\
&&
\!\!\!\!\!\!\!\!\!\!\!\!\!
D_{tH}\left\{ \frac{\tilde s\tilde t}{(1-\tilde s)^{2}}
+\frac{a_k+1-\tilde u}{\tilde t}\left[-\frac{\tilde s L_{2}(\tilde u)}{1-\tilde u}
%\right.\right. 
%\\
%&  & 
%\left.\left.
+\frac{(1-\tilde s)(a_k+\tilde s)
+a_k\tilde t}
{\tilde s\tilde u+a_k(1-\tilde s)}L_{3}(\tilde s,\tilde u)\right]\right\} 
\nonumber \\
&  & 
\!\!\!\!\!\!\!\!\!\!\!\!\!
+D_{u\ell}\left\{ \frac{\tilde s\tilde t}{(1-\tilde s)^{2}}
-\frac{\tilde u}{\tilde t(1-\tilde s)}\right.
-\frac{(a_k+1-\tilde u)\left[\tilde s\tilde t-
\tilde u(1-\tilde u)\right]}
{\tilde t^{2}(1-\tilde u)}L_{2}(\tilde u)
\label{eq:box2_cut1} \\
&  & 
\!\!\!\!\!\!\!\!\!\!\!\!\!
-\left.\left[\frac{1-3\tilde s-2a_k}{\tilde t}
+\frac{(1-\tilde s)(a_k+\tilde s)}{\tilde t^2}
%\nonumber \\
%&  & 
%\!\!\!\!\!\!\!\!\!\!\!\!\!
+\frac{a_k-\tilde s}{\tilde s}-\frac{a_k(a_k+\tilde s)}
{\tilde s^2\tilde u+a_k\tilde s(1-\tilde s)}
\right] L_{3}(\tilde s,\tilde u)\right\}.
\nonumber 
\end{eqnarray}

The corresponding contributions from $C_1^\prime$ cuts read
${{I'}^{N_i\bar\ell_j}_{A H}}^{(b)}(s,t,u)=$\\
${{I'}^{\bar\ell_j A}_{N_i H}}^{(b)}(s,t,u)=0$ and  
${{I'}^{N_i A}_{\ell_j H}}^{(b)}(s,t,u)={I'}^{(b)}(s,t,u)\equiv I'_{b1}(s,t,u)+
I'_{b2}(s,t,u)$, with
\begin{eqnarray}
&&
\!\!\!\!\!\!\!\!\!\!\!\!\!
I'_{b1}\left(s,t,u\right) = \nonumber \\
&&
\!\!\!\!\!\!\!\!\!\!\!\!\!
D_{tH}\left\{ -\frac{\tilde s\tilde u}{(1-\tilde s)^{2}}
+\frac{a_k+\tilde s}{\tilde u}L_{1}(\tilde s)
%\right.
%\label{eq:box1_cut2} \\
%&  & \left.
-\left[\frac{a_k+\tilde s}{\tilde u}+\frac{a_k(a_k+1-\tilde t)}
{\tilde s\tilde t+a_k(1-\tilde s)}\right]L_{3}(\tilde s,\tilde t)\right\} 
\nonumber \\
&&
\!\!\!\!\!\!\!\!\!\!\!\!\!
+D_{u\ell}\left[1+\frac{\tilde s\tilde t}{(1-\tilde s)^{2}}
-\frac{a_k+\tilde s}{\tilde s}L_{1}(\tilde s)
-\frac{(a_k +\tilde s)(a_k+1-\tilde t)}
{\tilde s\tilde t+a_k(1-\tilde s)}L_{3}(\tilde s,\tilde t)\right],
%\nonumber
\label{eq:box1_cut2} \\
%\end{eqnarray}
%\begin{eqnarray}
&&
\!\!\!\!\!\!\!\!\!\!\!\!\!
I'_{b2}\left(s,t,u\right)  =  \nonumber \\
&&
\!\!\!\!\!\!\!\!\!\!\!\!\!
D_{tH}\left\{ -\frac{\tilde s\tilde t}{(1-\tilde s)^{2}}
+\frac{a_k+1-\tilde u}{\tilde t}\biggl[L_{1}(\tilde s)
%\right.
%&  & \left.
\left.-\frac{\left(1-\tilde s\right)(a_k+\tilde s)
+a_k\tilde t}{\left[\tilde s\tilde u+a_k(1-\tilde s)\right]}
L_{3}(\tilde s,\tilde u)\right]\right\}  \label{eq:box2_cut2} \\
&&
\!\!\!\!\!\!\!\!\!\!\!\!\!
+D_{u\ell}\left\{ \frac{\tilde s}{\tilde t}
-\frac{1-\tilde s+\tilde s\tilde t}{(1-\tilde s)^{2}}
%\right.
%\nonumber \\
%&  & 
+\left[\frac{(1-\tilde u)^{2}(a_k+\tilde s)}
{\tilde s\tilde t^{2}}
+\frac{1-\tilde u}{\tilde t}-\frac{a_k+1-\tilde u}{\tilde t^{2}}\right]
L_{1}(\tilde s) \right.
\nonumber \\
&& 
\!\!\!\!\!\!\!\!\!\!\!\!\!
+\left.\left[\frac{1-3\tilde s-2a_k}{\tilde t}
+\frac{(1-\tilde s)(a_k+\tilde s)}{\tilde t^2}
%\nonumber \\
%&  & 
%\!\!\!\!\!\!\!\!\!\!\!\!\!
+\frac{a_k-\tilde s}{\tilde s}-\frac{a_k(a_k + \tilde s)}
{\tilde s^2\tilde u+a_k\tilde s(1-\tilde s)}
\right] L_{3}(\tilde s,\tilde u)\right\}.
\nonumber
\end{eqnarray}
From the Eqs.\eqref{eq:box1_cut1}--\eqref{eq:box2_cut2}, 
we see that for the scattering process $N_i A \to \ell_j H$ 
the contributions from $C_1$  and $C^\prime_1$ partially cancel each other
\footnote{By definition
${I^{12}_{34}}^{(b)}$, ${{I'}^{12}_{34}}^{(b)}$, and
${J^{12}_{34}}^{(b)}$  must be real. It can be easily
verified that in their expressions the imaginary contributions
from negative arguments in the logarithms in
Eq.~\eqref{eq:def_log} always cancel out.}.

The $C_2$  contributions to the box asymmetry are
${J^{N_i A}_{\ell_j H}}^{(b)}(s,t,u)=0$ and 
\begin{eqnarray}
&&
\!\!\!\!\!\!\!\!\!\!\!\!\!
{J^{N_i\bar\ell_j}_{A H}}^{(b)}\!\!\!\!(s,t,u) = \nonumber \\
&&
\!\!\!\!\!\!\!\!\!\!\!\!\! 
D_{sH}\left\{ \frac{\tilde u(a_k+1-\tilde s)}
{\tilde t(\tilde s-1)}L_{4}(\tilde s)
+\left[\frac{a_k+\tilde u}{\tilde t}+
\frac{a_k(a_k+1-\tilde s)}{\tilde s\tilde u+a_k(1-\tilde u)}
\right] L_{5}(\tilde u,\tilde s)\right\}  \\
&&
\!\!\!\!\!\!\!\!\!\!\!\!\!
+D_{t\ell}\left\{ \frac{(1-\tilde t)
(a_k-\tilde s)}{\tilde s^{2}}-\frac{(a_k+\tilde u)}
{\tilde s-1}L_{4}(\tilde s) 
%\right.
%\\
%&  & \left.
+\frac{(a_k+1-\tilde s)(a_k+\tilde u)}
{\tilde s\tilde u+a_k(1-\tilde u)}
L_{5}(\tilde u,\tilde s)\right\}, \nonumber \\
%\end{eqnarray}
%\begin{eqnarray}
&&
\!\!\!\!\!\!\!\!\!\!\!\!\!
{J^{\bar\ell_j A}_{N_i H}}^{(b)}\!\!\!\!
(s,t,u) =  D_{s\ell}\left\{ \frac{\tilde s-a_k}{\tilde t}
-\frac{(a_k-\tilde s)(a_k+\tilde u)}
{\tilde s\tilde u+a_k(1-\tilde u)}
L_{5}(\tilde u,\tilde s)\right.\nonumber \\
&&
\!\!\!\!\!\!\!\!\!\!\!\!\!
\left.+\frac{a_k+1-\tilde s}
{\tilde t^{2}}\left[\frac{\tilde t\tilde u-\tilde s(1-\tilde s)}
{1-\tilde s}L_{4}(\tilde s)
+(\tilde s-\tilde t)L_{5}(\tilde u,\tilde s)\right]
\right\}
\\
&&
\!\!\!\!\!\!\!\!\!\!\!\!\!
+D_{tH}\left\{
\frac{\tilde u(a_k+1-\tilde s)}{\tilde t(1-\tilde s)}L_{4}(\tilde s)
%\right.\\
%&  & \left.
-\frac{(a_k+1-\tilde s)
\left[(a_k+\tilde u)(1-\tilde u)+a_k \tilde t\right]}
{\tilde t\left[\tilde s\tilde u+a_k(1-\tilde u)\right]}
L_{4}(\tilde s)\right\}. 
\nonumber 
\end{eqnarray}
Finally the $C_2$  contributions from diagrams 3.(i) and 3.(j) read
\begin{eqnarray}
&& \!\!\!\! 
{J^{\bar\ell_j A}_{N_i H}}^{(\Delta)}(s,t,u) =- 
D_{s\ell}\frac{M_i^2(\tilde s-a_k)}{2\tilde s}
\left[\frac{\tilde s-a_k}{\tilde s}+D_{s\ell}M_i^2 (\tilde s+a_k)\right]
\nonumber \\
&&
\times \left[D_{tH}\tilde u + D_{s\ell}\tilde s(1-\tilde u)\right].
\end{eqnarray}

\begin{figure}
\includegraphics[width=0.5\textwidth]{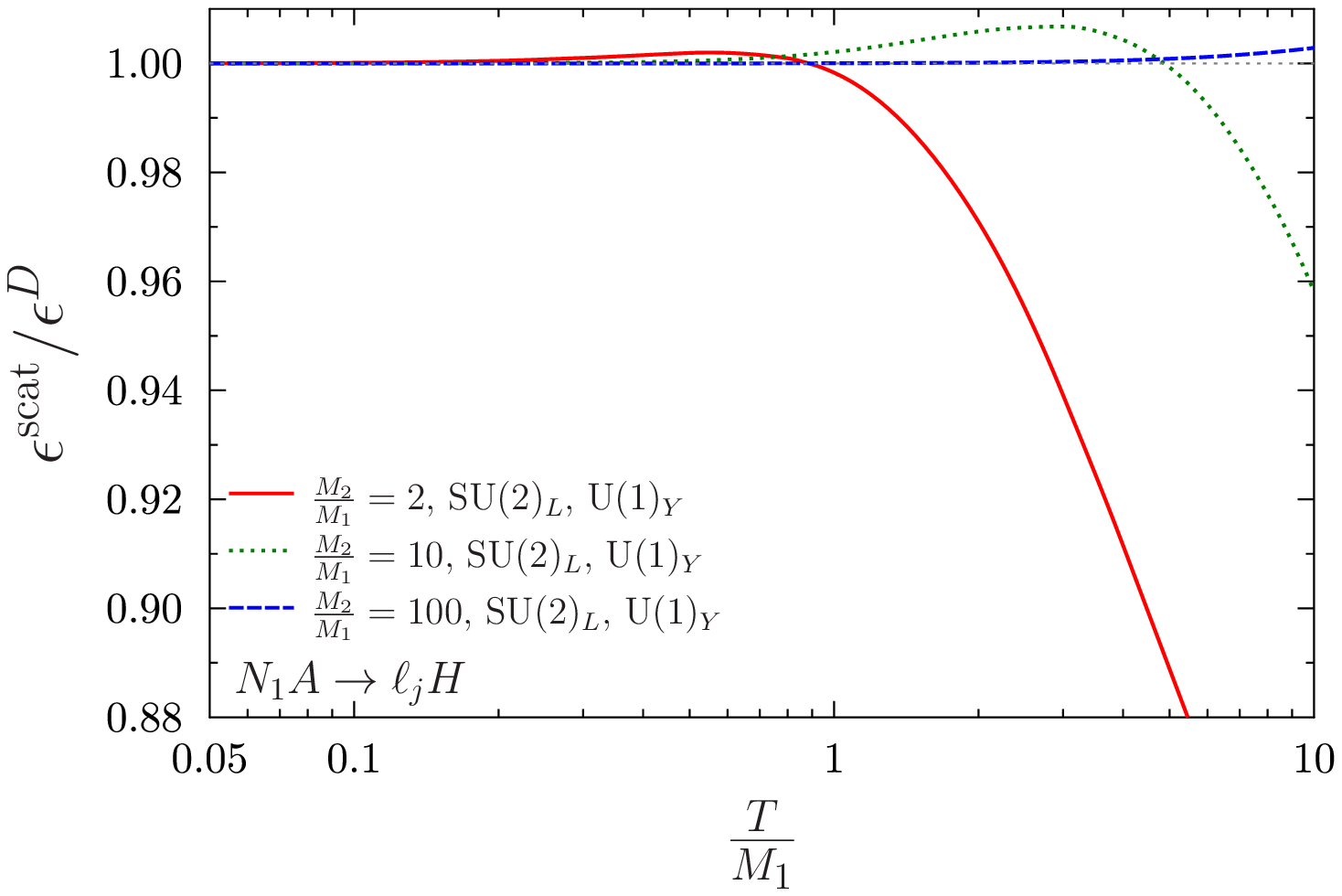}
\includegraphics[width=0.5\textwidth]{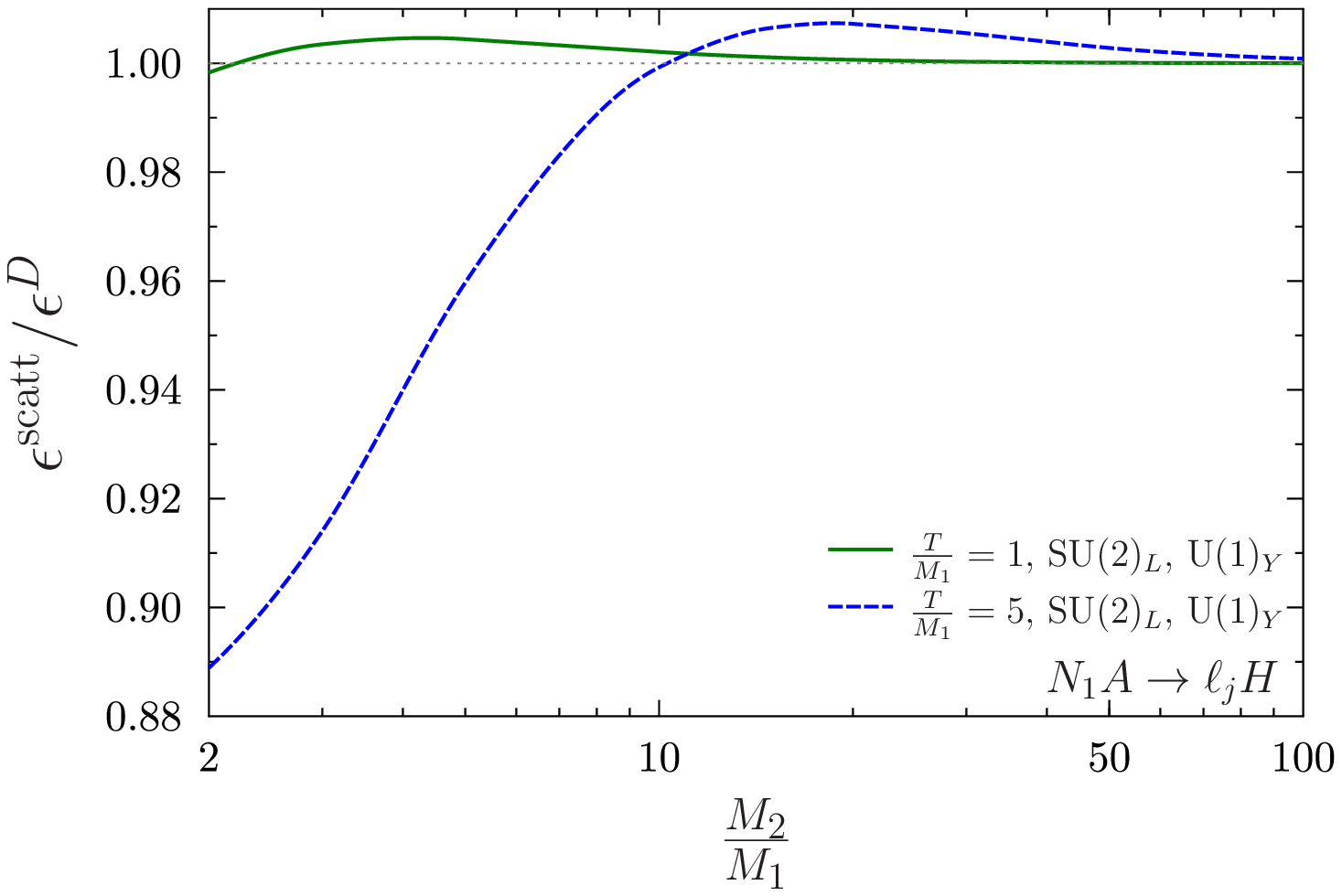}
\includegraphics[width=0.5\textwidth]{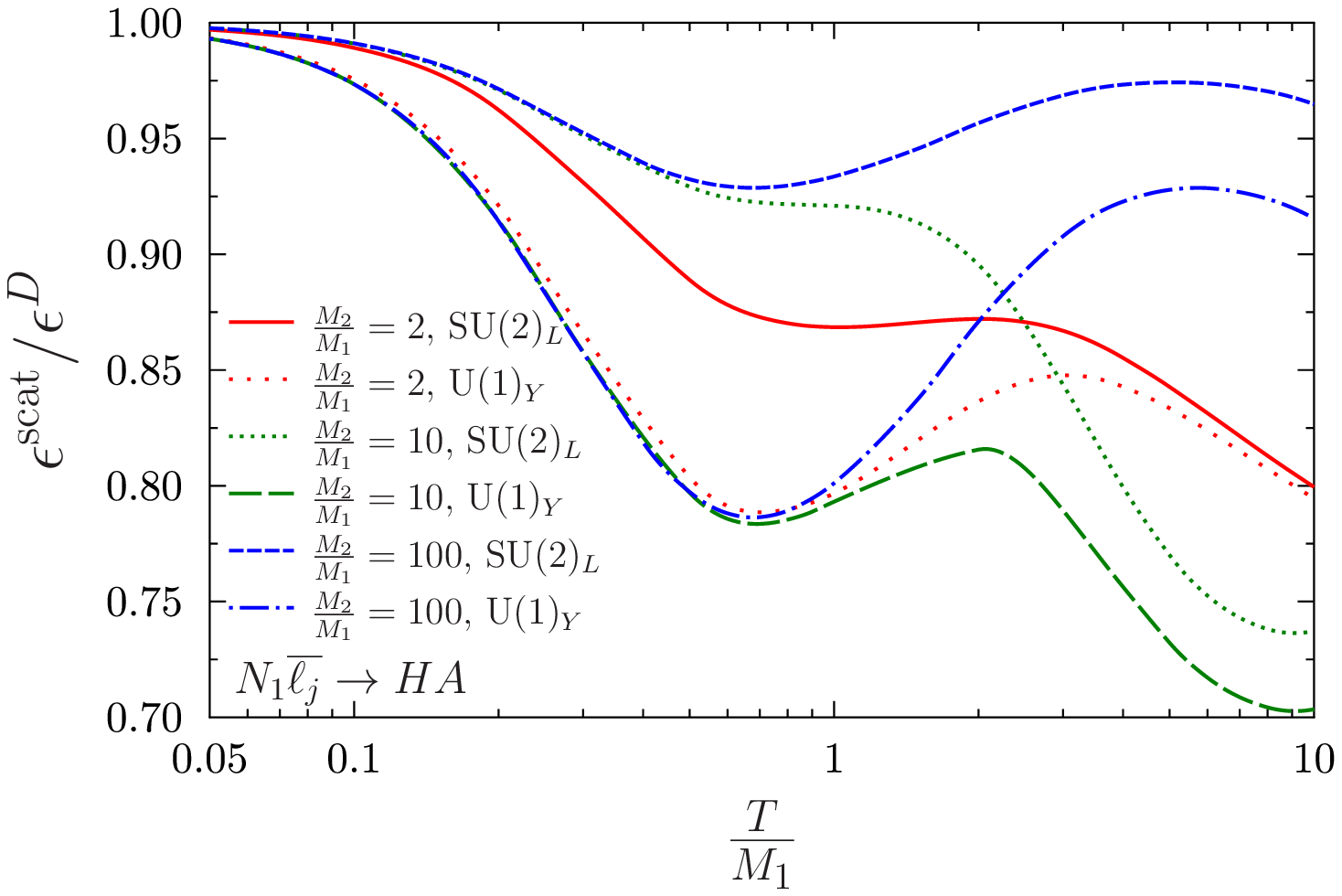}
\includegraphics[width=0.5\textwidth]{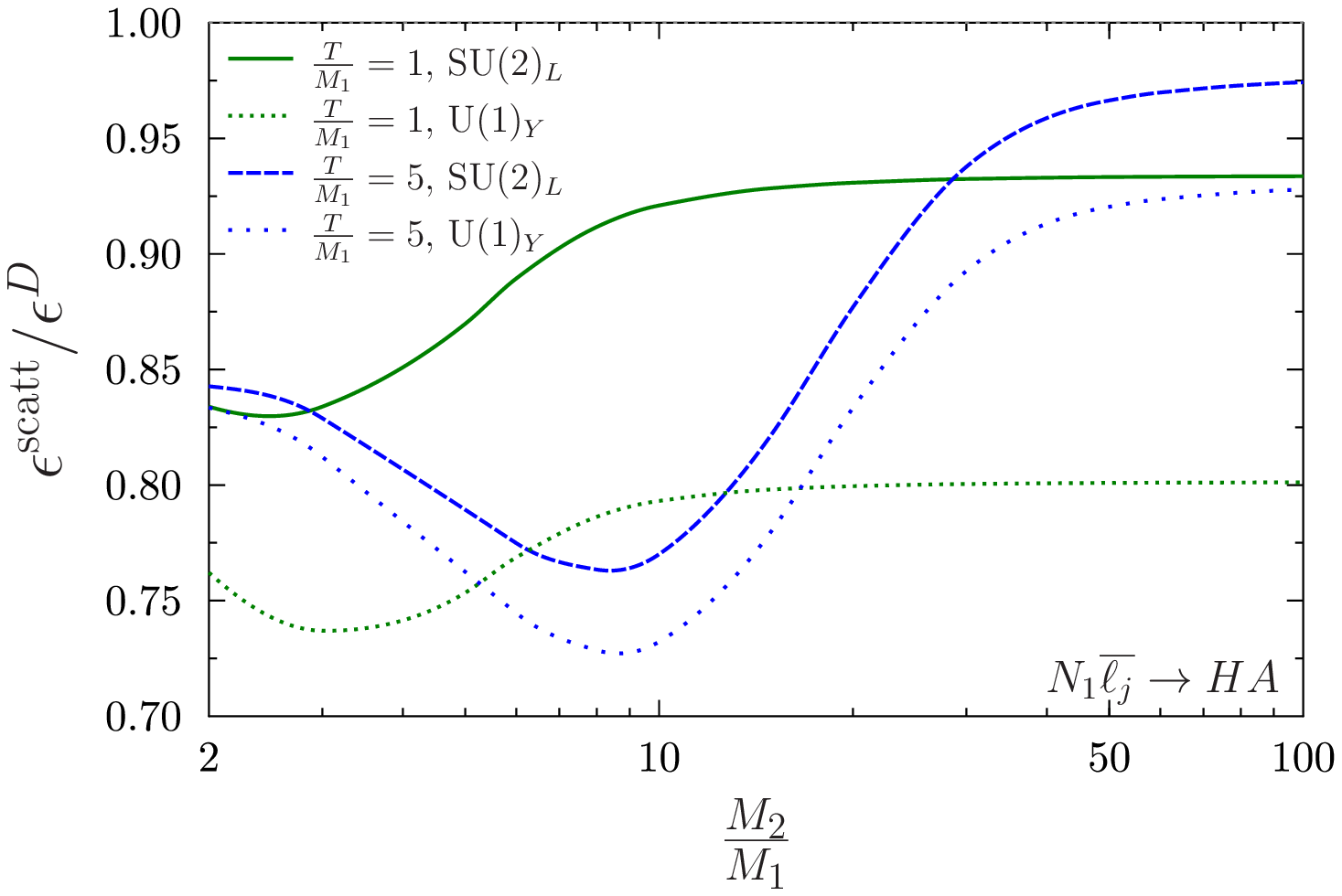}
\includegraphics[width=0.5\textwidth]{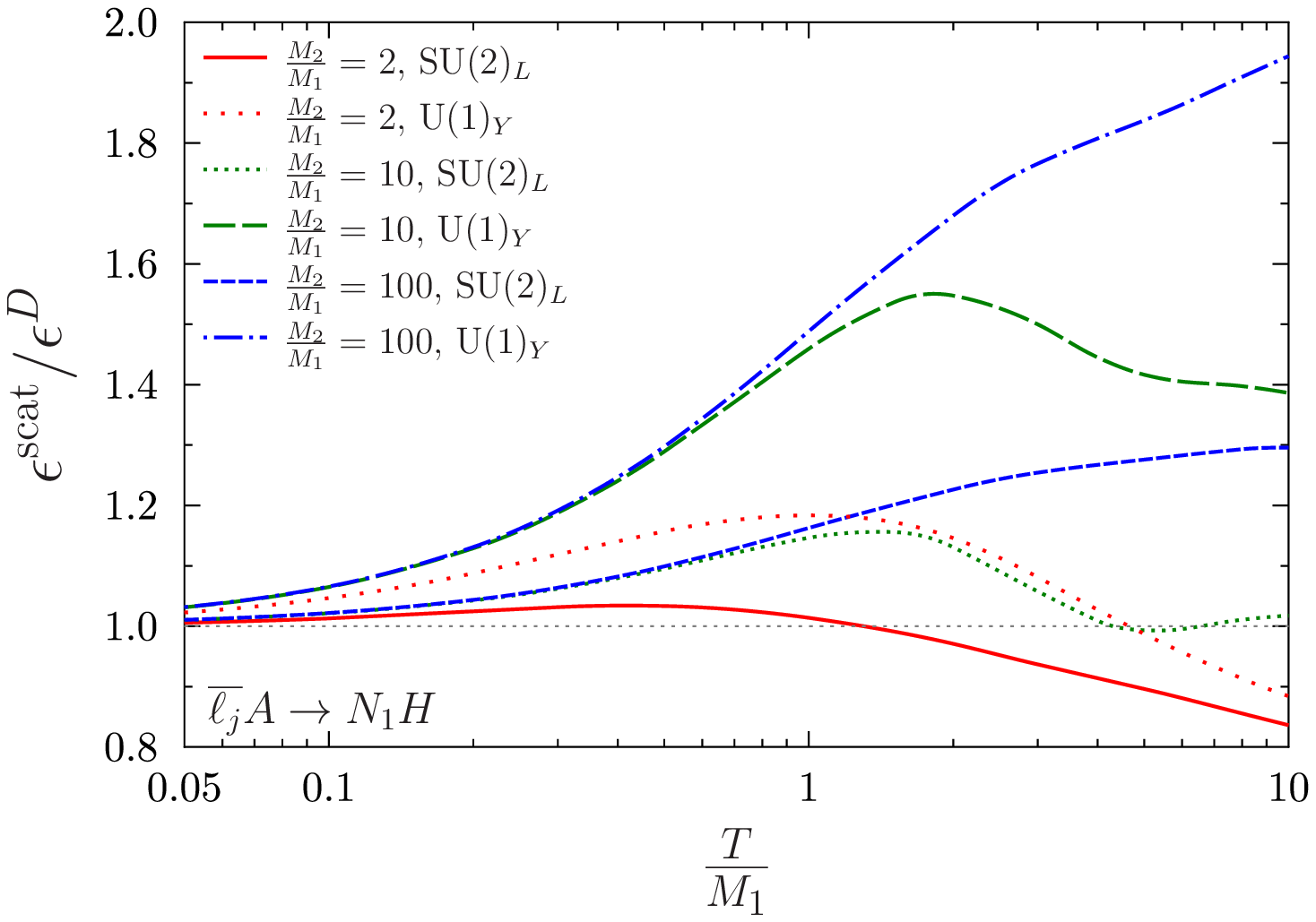}
\includegraphics[width=0.5\textwidth]{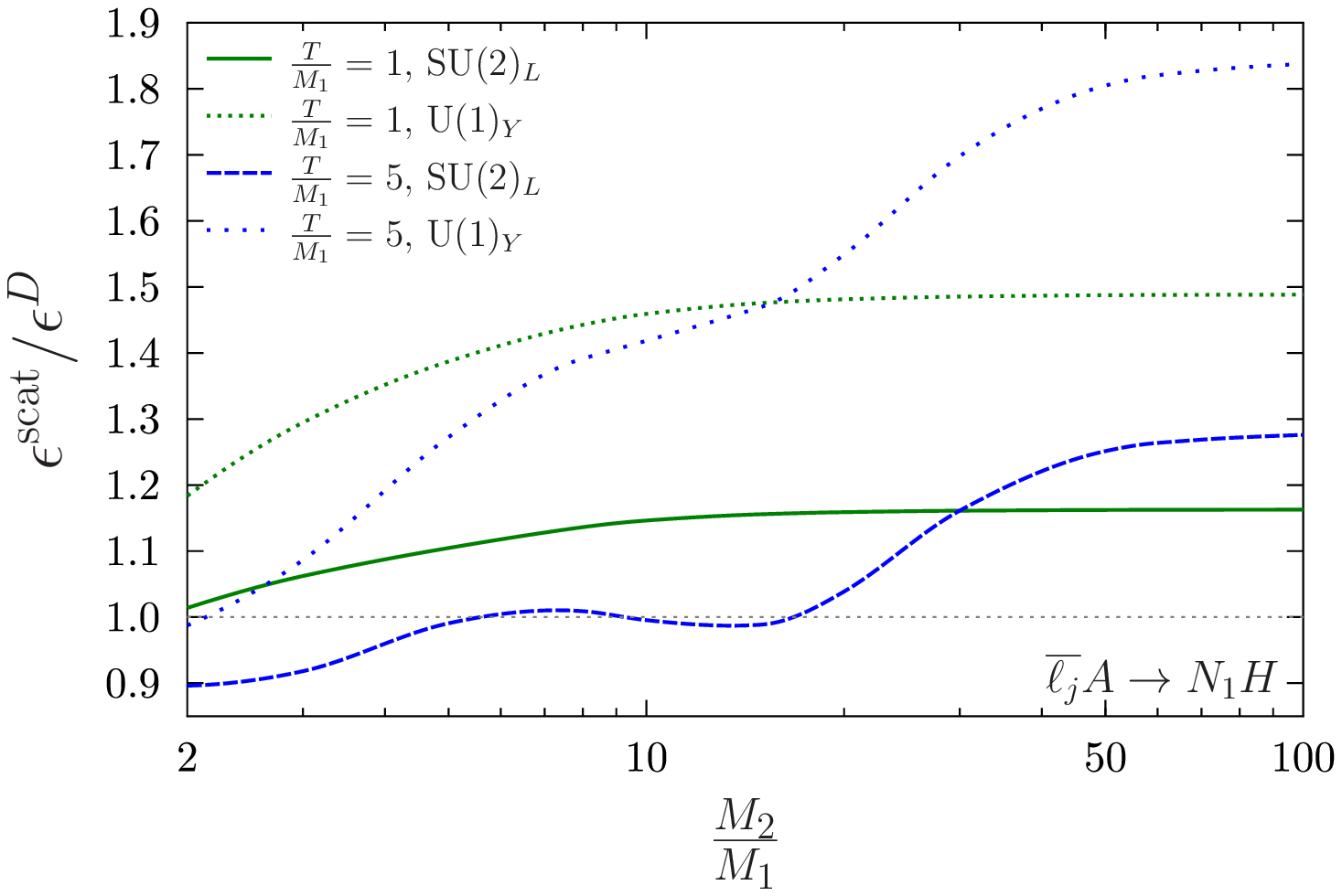}
\caption{Ratio between the L-violating asymmetry in scatterings
$\epsilon^{scat}\equiv \epsilon^{12}_{34}$ for the different processes
as labeled in the figure and the corresponding quantity for decays
$\epsilon^{D}\equiv \epsilon^{N_i}_{\ell_j H}$.  To illustrate we have
used $m_\ell = 0.18 T$ and $m_H= 0.34 T$.
The left panels show the ratio 
as  a function of $T/M_1$ for three different hierarchies  
$M_2/M_1=2,10,100$.
The right panels show the ratio as a function of $M_2/M_1$
for two values of $T/M_1=1,5$.}
\label{fig:cpasym}
\end{figure}

The ratios between the L-violating asymmetries in GBS and the
L-violating asymmetry in decays are shown in Fig.~\ref{fig:cpasym}
as a function of $T/M_1$ (left panels) and of $M_2/M_1$ (right
panels).  For simplicity we have considered the virtual effects of
only one neutrino species $N_k=N_2$ different from $N_i=N_1$. 
It is apparent
that for $N_{1}\overline{\ell}_{j} \to HA$ and $\overline{\ell}_{j}A \to
N_{1}H$
this ratio deviates from unity even for hierarchical neutrinos 
\footnote{Unlike the CP asymmetry in decays, the CP asymmetry 
in scatterings depends on the temperature, thus when
comparing both types of asymmetries at a given $T$, the RHNs
are considered to be hierarchical enough if $M_{k\neq 1} \gg M_1,T$.}.
In the relevant temperature range for leptogenesis $0.1 \lesssim T/M_1
\lesssim 10 $ the deviation can be of several tens of percent,
e.g. $50 \%$ for $\overline{\ell}_{j}B_Y \to N_{i}H$ at $T=M_1$.
Conversely for $N_{1}A \to \ell_{j}H$ the ratio tends to one for
hierarchical neutrinos, hence the asymmetry can be well approximated
by the decay one if $M_2/M_1\gg 10$, but corrections appear at high
temperatures for milder hierarchies.

These results can be easily understood analytically by
expanding to lowest order in  $M_i/M_k$ 
the expressions in 
Eqs.~\eqref{eq:vertex1}, \eqref{eq:vertexnew}, 
and  \eqref{eq:box1_cut1}--\eqref{eq:box2_cut2} which gives
\begin{eqnarray}
&& \!\!\!\!\!\!\!\!\!\!\!\!\!\!
I^{(v1)}(s,t,u)  =  -\frac{1}{2} \frac{I_0(s,t,u)}{M_k^2},
\label{eq:expansion1}
\\&&
\!\!\!\!\!\!\!\!\!\!\!\!\!\!
I^{(b)}(s,t,u)  =  
\frac{M_i^2}{M_k^2}\frac{
\tilde s \left\{ 
D_{u\ell}\tilde u\left[3\tilde t+\tilde u\left(1-\tilde s\right)\right]
-D_{tH}\tilde t\left[
\left(\tilde t-2\tilde u\right)+\tilde u\left(1-\tilde s\right)\right]
\right\}}{(\tilde t+\tilde u)^{3}},
\label{eq:expansion2}\\
&&\!\!\!\!\!\!\!\!\!\!\!\!\!\!
{I'}^{(b)}(s,t,u)  =  - I^{(b)}(s,t,u), \label{eq:expansion3}
\\
&&\!\!\!\!\!\!\!\!\!\!\!\!\!\!
K^{(v2)}(s,t,u)=2  I^{(b)}(s,t,u)=-K'^{(v2)}(s,t,u), \label{eq:expansion4}
\\
&&\!\!\!\!\!\!\!\!\!\!\!\!\!\!
\widetilde K^{(v2)}(s,t,u) \! = \! 
-\frac{M_i^2}{M_k^2} \frac{2 \tilde s\tilde u
\left\{D_{tH}\tilde t\left[3-2(\tilde t+\tilde u)\right]
+D_{u\ell}\left[(2\tilde t-\tilde u)-\tilde t^2+\tilde u^2\right] 
\right\}}{(\tilde t+\tilde u)^3}. \label{eq:expansion5}
\end{eqnarray}

Including these results in~Eq.\eqref{eq:verboxcp} and using 
Eqs.~\eqref{eq:tree1}--\eqref{eq:tree3} one finds that 
\begin{equation}
{\epsilon^{N_i A}_{\ell_j H}}^{(v1)}
={\epsilon^{N_i\bar\ell_j}_{A H}}^{(v1)}
={\epsilon^{\bar\ell_j A}_{N_i H}}^{(v1)}=
\frac{\displaystyle\sum_{k\neq i}
C_{jk}}{16\pi\left(Y_{N}Y_{N}^{\dagger}\right)_{ii}}\frac{M_{i}}{M_{k}}
={\epsilon^{N_i}_{\ell j H}}^{(v)}.
\label{eq:factorv}
\end{equation}
In the last equality we have used that the  vertex CP asymmetry 
from the decay  $N_{i}\to \ell_{j}H$
\begin{equation}
{\epsilon^{N_i}_{\ell j H}}^{(v)}=
-\frac{\displaystyle\sum_{k\neq i}C_{jk}}
{8\pi \left(Y_{N}Y_{N}^{\dagger}\right)_{ii}}
\sqrt{a_k}\left[1-(1+a_k)
\ln\frac{1+a_k}{a_k}
\right]\simeq 
\frac{\displaystyle\sum_{k\neq i}
C_{jk}\frac{M_{i}}{M_{k}}}{16\pi\left(Y_{N}Y_{N}^{\dagger}\right)_{ii}},
\end{equation}
where the last approximation holds for $M_k/M_i\gg 1$.
So, as in the case of TQS, both wave and
vertex contributions to the CP asymmetry in scatterings with SU(2)$_L$ 
gauge bosons  are equal to the corresponding contributions to the decay 
CP asymmetry  in the hierarchical limit ($M_k/M_i\gg 1$).

For the process $N_{i}A \to \ell_{j}H$
Eqs.~\eqref{eq:expansion2}--\eqref{eq:expansion4} imply that in the
hierarchical limit the CP symmetries from vertices (e) and (f) with
U(1)$_Y$ gauge bosons, and from boxes with either SU(2)$_L$ or
U(1)$_Y$ gauge bosons vanish,
\begin{equation}
{\epsilon^{N_i W_a}_{\ell_j H}}^{(b)}={\epsilon^{N_i B_Y}_{\ell_j H}}^{(b)}=
{\epsilon^{N_i B_Y}_{\ell_j H}}^{(v2)}=0.
\end{equation}
However for the other two processes we find that the L-violating  contributions
to the CP asymmetries satisfy
\begin{eqnarray} &&
{\epsilon^{N_i\bar\ell_j}_{W_a H}}^{(b)}\neq 0, 
\;\;\;\;\;\;\;\;\;\;\;\;\;\;\;\;\;\;\;\;\;\;\;\;\;
 {\epsilon^{\bar\ell_j W_a}_{N_i H}}^{(b)}\neq 0,  \nonumber\\
&& {\epsilon^{N_i\bar\ell_j}_{B_Y H}}^{(b)}
=\frac{1}{2}
{\epsilon^{N_i \bar\ell_j}_{B_Y H}}^{(v2)}\neq 0, 
\;\;\;\;\;\;
{\epsilon^{\bar\ell_j B_Y}_{N_i H}}^{(b)}
=\frac{1}{2}{\epsilon^{\bar\ell_j B_Y}_{N_i H}}^{(v2)}
\neq 0.
\label{eq:nofactorb}
\end{eqnarray}
We note that the vanishing of ${\epsilon^{N_i A}_{\ell_j H}}^{(b)}$
and ${\epsilon^{N_i B_Y}_{\ell_j H}}^{(v2)}$ 
at leading order in $M_i/M_k$ is due to the presence of the 
additional $C'_1$ cuts, which, as mentioned above, 
results in  a partial or total  cancellation  of the contributions from 
the $C_1$ cuts so that only terms of higher order in 
$M_i/M_k$  remain.

We have mentioned before that for each process the sum of amplitudes of 
wave diagrams is independently gauge invariant, as it is the 
sum of the amplitudes for the vertex ($v2$) diagrams, but 
the amplitudes of vertex ($v1$) and
box diagrams are not separately gauge invariant. However at the leading 
order in  $M_i/M_k$, the box and vertex ($v1$) diagrams are independently
gauge invariant, thus the results in Eq.\eqref{eq:factorv} 
and  Eq.\eqref{eq:nofactorb} are also gauge independent.

Summarizing, we find that even for very hierarchical heavy RHNs
\begin{equation}
\epsilon^{N_i A}_{\ell_j H}={\epsilon^{N_i}_{\ell j H}} \neq
\epsilon^{N_i\bar\ell_j}_{A H}\neq
\epsilon^{\bar\ell_j A}_{N_i H},
\end{equation}
where ${\epsilon^{N_i}_{\ell j H}} = {\epsilon^{N_i}_{\ell j H}}^{(w)} + {\epsilon^{N_i}_{\ell j H}}^{(v)}$. 
Hence, we conclude that the CP asymmetry 
in the scattering $N_{i}A\to \ell_{j}H$ factorizes in the hierarchical limit. 
However, this is not the case for the scatterings 
$N_i\overline{\ell}_j\to HA$ and $\overline{\ell}_jA \to N_i H$.

Also, as mentioned above, in the processes
$N_{i}\overline{\ell}_{j} \to H B_Y$ and $\overline{\ell}_{j} B_Y \to
N_{i}H$ there are new L-conserving vertex contributions to the CP
asymmetry. They imply that in the hierarchical
limit the L-conserving CP asymmetry in these scatterings is not equal
to the L-conserving CP asymmetry in decays. Note that although the
L-conserving asymmetries are always suppressed by a higher power of
$M_i/M_k$ compared to the L-violating ones, they are the only source
of CP violation in models with conservation of
L~\cite{aristizabal09b,ourL,antusch09}.

Finally, a few comments about the impact of our results for the
determination of the baryon asymmetry are in order. Although the CP
asymmetry in GBS has more interesting features
compared to the one in TQS, we have found
that still the CP asymmetry per scattering is $(0.5-2) \times$ the CP
asymmetry per decay. Therefore the generation of CP asymmetry through
GBS is relevant at high temperatures ($T \gtrsim
M_i$) when the scattering rates are larger than the decay rates. 
Hence it can have effects in models for leptogenesis that are sensible
to the high temperature regime. This includes standard type I
leptogenesis in the weak washout regime and low energy models in which
the sphaleron processes freeze out before the RHNs have
decayed~\cite{pilaftsis08,ourL}. 

\vskip 0.3cm 
%\ack
We thank E. Nardi, N. Rius, Y. Nir,  
J. Salvado, A. Gadde and I. Sung 
for useful discussions and suggestions.
This work is supported by USA-NSF grant PHY-0653342, by
Spanish grants from MICINN 2007-66665-C02-01, the INFN-MICINN
agreement program ACI2009-1038, consolider-ingenio 2010 program
CSD-2008-0037 and by CUR Generalitat de Catalunya grant 2009SGR502.

\end{document}